\documentstyle[aps,psfig,prb,preprint]{revtex}
\begin{document}
\title{Semiclassical theory of vibrational energy relaxation}
\author{Robert Karrlein and Hermann Grabert}
\address{Fakult\"{a}t f\"{u}r Physik der Albert--Ludwigs--Universit\"{a}t,
 Hermann-Herder-Str.~3,\\ D-79104 Freiburg, Germany}
\date{November 12, 1997}
\maketitle
\begin{abstract}
A theory of vibrational energy relaxation based on a semiclassical
treatment of the
quantum master equation is presented. Using new results on the
semiclassical expansion of dipole matrix elements, we show that in the
classical limit the master equation reduces to the Zwanzig energy
diffusion equation. The leading quantum corrections are determined and
discussed for the harmonic and Morse potentials. 
\end{abstract}
\newpage
\section{Introduction}
Vibrational energy relaxation (VER) is of fundamental importance for
many chemical reactions in solution \cite{reviews,owrutsky}. VER studies
in molecules like ${\rm I}_2, {\rm HgI}, {\rm O}_2$, and 
${\rm N_2}$ in various solvents by means of pulsed laser techniques
cover a range of relaxation times from picoseconds to seconds
\cite{owrutsky}. A detailed theoretical description is often
complicated by the fact that various energy relaxation processes
contribute. While in the high frequency regime a quantum mechanical
treatment, e.~g., on the basis of the isolated binary collision model
is natural, the low frequency regime has for the most part been
treated in terms of classical theories such as the Landau-Teller
model, the generalized Langevin equation, or by molecular dynamics
\cite{classical,zwanzig,tuckerman}. These approaches are justified for
high temperatures and heavier molecules.

However, in some of the above mentioned molecules in solution as well
as for adsorbates on surfaces, molecules in cryogenic matrices, or
molecular crystals, purely classical theories are not sufficient and
quantum effects cannot be disregarded. This is
also due to the fact that experiments have to be performed at low
temperatures where only few vibrational levels are occupied. For these
systems, VER was mostly treated on the basis of Fermi's golden rule
and the quantum master equation \cite{fermi,bader}. 

One of the questions arising in this context is the connection between
the quantum mechanical theory of VER and classical energy
relaxation. To address this problem, we give here a semiclassical
analysis of the quantum master equation. The Zwanzig energy diffusion
equation \cite{zwanzig} is recovered in the classical limit and the dominant
quantum corrections are determined explicitly.

The paper is organized as follows. In section II we
introduce the microscopic model which is a nonlinear oscillator
coupled to a bath of harmonic oscillators. This model is known to lead
in the classical limit to the generalized Langevin equation. Molecular
dynamics simulation suggest that the model can also be used for rather
complicated relaxation processes, such as e.g., a binary molecule in a
Lennard-Jones fluid, if the parameters of the harmonic oscillator bath
are chosen appropriately \cite{tuckerman}. 
We briefly discuss the classical limit of
the model, and its weak-coupling limit, which leads to the Zwanzig energy
diffusion equation. Further, the fully quantum mechanical treatment of
VER on the basis of a master equation is given. 

Section III is devoted to the semiclassical expansion of the
master equation. This requires, apart from known expansions of wave
functions and energy levels, explicit results on semiclassical dipole
matrix elements. On the basis of these results we then show that the
master equation reduces to the classical energy diffusion equation in
the limit $\hbar\to 0$, and determine the leading quantum
corrections. Section IV contains explicit results for the cases of the
harmonic and Morse potentials, and in section V we present our
conclusions.
\section{The model and formulation of the problem}
Vibrational motion of molecules can be modeled as the motion of a
particle in an anharmonic potential well $V(q)$. To account for
dissipation we couple the particle to a bath of harmonic oscillators
\cite{revCLmodel}. Of course, the harmonic oscillator bath is a crude
simplification of the real interaction of molecular vibrations with
the solvent. However, inasmuch as the back-action of the solvent can
be treated within linear response theory, the bath can be modeled in
terms of effective bath oscillators. In fact, this simple
model describes the classical dynamics of a binary molecule in a
Lennard-Jones fluid quite accurately \cite{tuckerman}. For the problem
of energy relaxation we expect the harmonic oscillator bath to be a
reasonable starting point. With these premises, the Hamiltonian $H$ is
composed of the oscillator part
\begin{equation}
H_{\rm S}={p^2\over 2M}+V(q),\label{HS}
\end{equation}
the bath Hamiltonian
\begin{equation}
H_{\rm B}=\sum_{n=1}^N {p_n^2\over 2 m_n}+{m_n\over 2} \omega_n^2 x_n^2,
\end{equation}
and an interaction part
\begin{equation}
H_{\rm SB}=-q \sum_{n=1}^N c_n x_n +q^2 \sum_{n=1}^N {c_n^2\over 2 m_n
\omega_n^2}.\label{SB}
\end{equation}
It is well known \cite{revCLmodel} that the associated classical 
equation of motion is the generalized Langevin equation
\begin{equation}
M\ddot{q}(t)+V'(q(t))+M \int_0^tds\gamma(t-s)\dot{q}(s)=\xi(t)
\end{equation}
where the prime denotes differentiation with respect to the argument, and
\begin{equation}
\gamma(t)={1\over M}\sum_{n=1}^N{c_n^2\over
m_n\omega_n^2}\cos(\omega_n t)
\end{equation}
is the damping kernel. Finally, $\xi(t)$ is a stochastic force having the
properties 
\begin{equation}
\langle\xi(t)\rangle=0
\end{equation}
and
\begin{equation}
\langle \xi(t)\xi(0)\rangle=Mk_BT\gamma(t).
\end{equation}
We will be interested in the motion at weak damping. When
$\omega_0$ is the frequency of small undamped oscillations about the
minimum of the potential well, a typical damping strength is given by
\begin{equation}
\gamma_c=\int_0^\infty dt\, \gamma(t)\cos(\omega_0 t).
\label{moment}
\end{equation}
Further, we may introduce a typical memory delay time
\begin{equation}
\tau_c={\int_0^\infty dt\, t\gamma(t)\cos(\omega_0 t)\over\int_0^\infty dt\, \gamma(t)\cos(\omega_0 t)}.
\end{equation}
The region of damping parameters examined in this article is then
characterized by
\begin{equation}
\gamma_c\ll\omega_0,1/\tau_c.\label{weak}
\end{equation}

Let us briefly recall the {\em classical treatment} of this weak damping limit.
Under the conditions stated, the unperturbed energy 
\begin{equation}
E={M\over 2}\dot{q}^2+V(q)
\end{equation}
of the particle will be an almost conserved quantity, and it is
advantageous to rewrite the classical Langevin equation in terms of energy and
phase variables.

The unperturbed trajectory $q(E,t)$ of energy $E$ satisfying
\begin{equation}
M\ddot{q}+V'(q)=0
\end{equation}
may be written as a Fourier series
\begin{equation}
q(E,t)=\sum_{l=-\infty}^\infty Q_l(E) e^{i l \Phi(t)}
\label{fourier}
\end{equation}
where
\begin{equation}
\Phi(t)=\omega(E) t.
\end{equation}
Here $2\pi/\omega(E)$ is the oscillation period of a trajectory of
energy $E$ in the potential well. The phase will be chosen such that
for $\Phi=0$ the trajectory is at a turning point. Then the Fourier
coefficients are real and satisfy 
\begin{equation}
Q_{-l}(E)=Q_l(E).
\end{equation}
Now, relation (\ref{fourier}) and
\begin{equation}
\dot{q}(E,t)=\sum_{l=-\infty}^\infty il\omega(E) Q_l(E) e^{i l \Phi(t)}
\end{equation}
define a mapping from the variables $q,\dot{q}$ to the variables
$E,\Phi$. By means of this transformation, the Langevin equation may
be written in terms of energy and phase 
\cite{zwanzig,kramers,linkwitz}. 
These transformed equations may be solved perturbatively to second
order in the coupling to the bath. Within the time window 
\begin{equation}
1/\omega_0,\tau_c\ll t\ll 1/\gamma_c
\end{equation}
the solution can be matched with the solution of a
Fokker-Planck equation for the energy and phase distribution function
\cite{linkwitz}. The structure of the resulting Fokker-Planck process
allows one to average over the phase variable yielding the
one-dimensional Fokker-Planck equation
\begin{equation}
\dot{P}(E)={\partial\over\partial E}\Lambda(E)\left(1+k_B
T{\partial\over\partial E}\right){\omega(E)\over 2\pi}P(E).\label{zwanzig}
\end{equation}
where
\begin{equation}
\Lambda(E)=2\pi M\omega(E)\sum_{l=-\infty}^\infty l^2 Q_l^2(E)\hat{\gamma}(i l\omega(E))
\label{Lambda}
\end{equation}
is the energy relaxation coefficient. Here 
\begin{equation}
\hat{\gamma}(z)=\int_0^\infty dt e^{-zt}\gamma(t)
\end{equation}
is the Laplace transform of the damping kernel.
Eq.~(\ref{zwanzig}) is the energy diffusion equation first derived by
Zwanzig \cite{zwanzig}. 

We now turn to the problem of energy relaxation in a weakly damped
{\em quantum system}. Introducing the energy eigenstates of the unperturbed
system 
\begin{equation}
H_{\rm S}|n\rangle=E_n|n\rangle,\quad n=0,1,2,\dots.
\end{equation}
we start from the Pauli master equation
\begin{equation}
\dot{p}_n(t)=\sum_{m=0}^\infty\left[W_{n,m}p_m(t)-W_{m,n}p_n(t)\right]
\label{master}
\end{equation}
where $p_n(t)$ is the probability to find the system in state $n$
at time $t$. The transition rate from state $m$ to $n$ may be
calculated by Fermi's golden rule
\begin{equation}
W_{n,m}={1\over\hbar^2}\int_{-\infty}^\infty dt
{\rm Tr}_{\rm B}\left[\langle n|H_{\rm SB}(t)|m\rangle
\langle m|H_{\rm SB}|n\rangle\rho_{\rm B}^{\rm eq}\right].
\end{equation}
Here ${\rm Tr}_{\rm B}$ denotes the trace over the bath
states, $\rho_{\rm B}^{\rm eq}=e^{-\beta H_{\rm B}}/{\rm Tr}_{\rm B}
[e^{-\beta H_{\rm B}}]$ is the equilibrium bath density matrix, and
$H_{\rm SB}(t)=e^{i(H_{\rm S}+H_{\rm B})t/\hbar}H_{\rm SB}
e^{-i(H_{\rm S}+H_{\rm B})t/\hbar}$.
Using Eq.~(\ref{SB}), we obtain
\begin{equation}
W_{n,m}={1\over\hbar^2}\langle n|q|m\rangle^2\sum_{k,l=1}^N c_kc_l
\int_{-\infty}^\infty dt e^{i(E_n-E_m)t/\hbar}{\rm Tr}_{\rm B}
\left[x_k(t)x_l\rho_{\rm B}^{\rm eq}\right].
\end{equation}
In the weak damping limit considered here, the so-called counter-term,
that is the second term of $H_{\rm SB}$ in Eq.~(\ref{SB}), may be
disregarded. The master equation (\ref{master}) is appropriate for
weak damping satisfying condition (\ref{weak}). In addition, for the quantum
system a further frequency scale set by the lowest Matsubara frequency
$\nu\equiv 2\pi k_BT/\hbar$ becomes relevant. As can be seen from the
derivation of the Pauli master equation, bath correlation functions
are assumed to decay on a time scale much shorter than $1/\gamma_c$
(see for instance Ref. \onlinecite{louisell}). For low temperatures
where $\nu<1/\tau_c$ bath correlations decay on the time scale
$1/\nu$. Hence, in addition to Eq.~(\ref{weak}) we have to require
\begin{equation}
\gamma_c\ll\nu
\end{equation}
which is, however, usually obeyed for systems of relevance in chemical
physics.

Due to the linearity of the bath, the transition
rate may be evaluated exactly yielding
\begin{equation}
W_{n,m}={1\over\hbar^2}\langle n|q|m\rangle^2 D(E_n-E_m).
\label{transi}
\end{equation}
Here
\begin{equation}
D(E)=2 M E \bar{n}(E){\rm Re}\, \hat{\gamma}(i|E|/\hbar)\label{SE}
\end{equation}
is an effective, occupation weighted spectral density of the bath with
the Bose occupation function
\begin{equation}
\bar{n}(E)={1\over e^{\beta E}-1}.\label{barn}
\end{equation}
Note that the coupling to the bath is completely described in terms of
the Laplace transform of the damping kernel. Since the detailed
balance condition
\begin{equation}
W_{m,n}=e^{\beta(E_n-E_m)}W_{n,m}
\label{detail}
\end{equation}
is clearly obeyed, it is readily seen that Eq.~(\ref{master}) has the
stationary solution 
\begin{equation}
p_n={1\over Z}e^{-\beta E_n}
\label{stationary}
\end{equation}
where
\begin{equation}
Z=\sum_{n=0}^\infty e^{-\beta E_n}
\end{equation}
is the quantum mechanical partition function.
\section{Semiclassical expansion of the Pauli master equation}
We now want to regain the Zwanzig energy diffusion equation from the
Pauli master equation and determine the leading quantum corrections.
The results of this somewhat lengthy calculation are summarized at the
beginning of section IV. We start by introducing the probability
distribution function 
\begin{equation}
P(E):=\sum_{n=0}^\infty \delta(E-E_n)p_n
\label{def}
\end{equation}
by means of which expectation values are simply calculated as
\begin{equation}
\langle f(E)\rangle=\int_{-\infty}^\infty dEP(E)f(E)
\label{expectation}
\end{equation}
for arbitrary $f(E)$. In terms of the function $N(E)$ which counts the
number of energy levels below energy $E$
\begin{equation}
N(E)=\sum_{n=0}^\infty \Theta(E-E_n),
\end{equation}
the probability $P(E)$ can be written
\begin{equation}
P(E)=\rho(E)p_{N(E)}
\end{equation}
where
\begin{equation}
\rho(E)={\partial N(E)\over \partial E}=\sum_{n=0}^\infty \delta(E-E_n)
\label{defrho}
\end{equation}
is the level density. Now, combining Eqs.~(\ref{master}) and (\ref{def})
we find
\begin{eqnarray}
\dot{P}(E)&=&\sum_{n,m=0}^\infty [\delta(E-E_m)-\delta(E-E_n)] W_{m,n}
\, p_n \nonumber\\ 
&=&\sum_{n=0}^\infty\sum_{l=-n}^\infty 
[\delta(E-E_{n+l})-\delta(E-E_n)]W_{n+l,n}\, p_n \label{zwischen}.
\end{eqnarray}
This equation will be evaluated in the semiclassical limit in the
sequel.
\subsection{Semiclassical energies}
To proceed we need to determine the semiclassical energy eigenvalues
and dipole matrix elements. We start from the semiclassical (WKB)
expansion of the wave function
\begin{equation}
\langle q|E\rangle=\sqrt{2 M\over \pi C(E)}e^{iS(E,q)/\hbar}
\end{equation}
where $C(E)$ is a normalization factor.
Inserting this ansatz into the Schr\"{o}dinger equation and expanding
in powers of $\hbar$, one arrives at the usual eikonal expansion
\begin{equation}
S(E,q)=S_0(E,q)-i\hbar S_1(E,q)-\hbar^2 S_2(E,q)+i\hbar^3 S_3(E,q)+\dots
\end{equation}
where the $S_i(E,q)$ can be calculated recursively with the result
\cite{bender}
\begin{eqnarray}
{\partial\over\partial q} S_0(E,q)&=&p(E,q)
=\sqrt{2M[E-V(q)]},\nonumber\\
{\partial\over\partial q} S_1(E,q)&=&-{1\over 2}
{\partial\over\partial q} \ln p(E,q),\nonumber\\
{\partial\over\partial q} S_2(E,q)&=&
{1\over 4 p(E,q)^2}{\partial^2\over\partial q^2} p(E,q)
-{3\over 8 p(E,q)^3}\left[{\partial\over\partial q} p(E,q)\right]^2,\\
{\partial\over\partial q} S_3(E,q)&=&{1\over 2}
{\partial\over\partial q}{1\over p(E,q)}{\partial\over\partial q} S_2(E,q).
\nonumber
\end{eqnarray}
To second order in $\hbar$ the wave function is then given by
\begin{eqnarray}
\langle q| E\rangle^{(\pm)} &=&\sqrt{2M\over\pi C(E)p(E,q)}
\exp\Biggl\{\pm{i\over\hbar} S_0(E,q)\pm{\hbar\over i} S_2(E,q)\nonumber\\
&&-{\hbar^2\over 2 p(E,q)} {\partial S_2(E,q)\over\partial q}+{\cal O}
(\hbar^3)\Biggr\}\label{WKBwave}.
\end{eqnarray}
The solution to the two turning point problem we are facing
here is a linear combination of these right and left traveling waves.
Details of the analytic structure of these functions are discussed
in the literature \cite{bender}. Here we only mention that the
quantization condition is obtained as
\begin{equation}
{i\over\hbar}\oint dq {\partial S(E,q)\over\partial q}
=2\pi i n,\ n=0,1,2,\dots.
\end{equation}
by requiring the single-valuedness of the wave function.
The integration contour is chosen to enclose the two turning points
at energy $E$. Defining the function
\begin{equation}
N_{\rm sc}(E):={1\over 2\pi\hbar}\oint dq {\partial S(E,q)\over\partial q},
\label{defNsc}
\end{equation}
this quantization condition may be written as
\begin{equation}
N_{\rm sc}(E)=n,\ n=0,1,2,\dots.\label{quant}
\end{equation}
The semiclassical energy eigenvalues
$E_{\rm sc}(n)$ ($n=0,1,2,\dots$) are then obtained by inverting
Eq.~(\ref{quant}). Now, the semiclassical expansion of
Eq.~(\ref{defNsc}) gives \cite{bender}
\begin{equation}
N_{\rm sc}(E)={1\over\hbar}J_0(E)-{1\over 2}-\hbar J_2(E)+{\cal O}(\hbar^3)
\label{wkb}
\end{equation}
where
\begin{equation}
J_{2 k}(E)={1\over 2\pi}\oint dq {\partial S_{2k}(E,q)\over\partial q}.
\label{defIk}
\end{equation}
Here $J_0(E)$ is the action variable with the property
\begin{equation}
{\partial J_0(E)\over\partial E}={1\over\omega(E)}\label{J0strich}.
\end{equation}
Truncation of the series (\ref{wkb}) at order
$\hbar$ reproduces the Bohr-Sommerfeld quantization
condition. Note that it is necessary to distinguish the semiclassical
quantity $N_{\rm sc}(E)$ from the exact $N(E)$ which is a staircase
function. For potentials where the series expansion (\ref{wkb}) terminates
at some finite order, $N_{\rm sc}(E)$ smoothly interpolates between
the sharp steps of $N(E)$. Similarly, the exact quantum mechanical
density $\rho(E)$ has to be distinguished from the smooth
semiclassical level density 
\begin{equation}
\rho_{\rm sc}(E):={\partial N_{\rm sc}(E)\over\partial E}
={1\over\hbar\omega(E)}-\hbar J_2'(E)+{\cal O}(\hbar^3).
\label{coarse}
\end{equation}
Accordingly, we define a generalized oscillation frequency containing
quantum corrections by
\begin{equation}
\Omega(E):={1\over \hbar\rho_{\rm sc}(E)}=
\omega(E)+\hbar^2 J_2'(E)\omega(E)^2+{\cal O}(\hbar^4)\label{defom}.
\end{equation}
Another quantity useful below is the energy difference
between nearby levels. Defining
\begin{equation}
\hbar A_l(E):=E_{\rm sc}(N_{\rm sc}(E)+l)-E,
\label{defAl}
\end{equation}
we find with the help of Eq.~(\ref{wkb})
\begin{equation}
\hbar A_l(E)=\hbar A_l^{(1)}(E)+{\hbar^2\over 2} A_l^{(2)}(E)
+{\hbar^3\over 3!}A_l^{(3)}(E)+{\hbar^4\over 4!}A_l^{(4)}(E)
+{\cal O}(\hbar^5),
\label{semicl}
\end{equation}
where
\begin{eqnarray}
A_l^{(1)}&=&l\omega,\nonumber\\
A_l^{(2)}&=&{l^2\over 2}\left[\omega^2\right]',\nonumber\\
A_l^{(3)}&=&6 l\omega^2 J_2'+{l^3\over 2}\omega\left[\omega^2\right]'',\\
A_l^{(4)}&=&12l^2\left[\omega^3J_2'\right]'+{l^4\over 2}\omega
\left\{\omega\left[\omega^2\right]''\right\}'.\nonumber
\end{eqnarray}
Here we have suppressed the explicit energy dependence for clarity.
\subsection{Semiclassical matrix elements}
As a first step to a calculation of semiclassical matrix elements the
normalization factor $C(E)$ needs to be determined. Extending previous
work on semiclassical matrix elements \cite{more}, one
finds  to second order in $\hbar$
\begin{equation}
C(E)\approx{M\over 2\pi}\oint{dq\over p(E,q)}\left[1-\hbar^2 {1\over p(E,q)}
{\partial S_2(E,q)\over\partial q}\right]={1\over\omega(E)}+h^2 C^{(2)}(E)
\label{Cerg}
\end{equation}
where the last equality defines $C^{(2)}(E)$. Below we will need
semiclassical expressions for the dipole matrix element
\begin{equation}
Q_l(E):=\langle E+\hbar A_l(E)|q|E\rangle\label{Ql}.
\end{equation}
It is, however, more convenient to evaluate the momentum matrix
element 
\begin{equation}
P_l(E):=\langle E+\hbar A_l(E)|p|E\rangle\label{Pl}
\end{equation}
from which the dipole matrix element can be obtained by means of the relation
\begin{equation}
P_l(E)=i M A_l(E)Q_l(E)\label{PQ}
\end{equation}
which is a consequence of $[H,q]=\hbar p/iM$. Combining 
Eqs.~(\ref{Ql}), (\ref{Pl}), (\ref{PQ}), and (\ref{WKBwave}) one finds
along these lines
\begin{eqnarray}
Q_l(E)&=&{1\over i M A_l(E)}
\int_{-\infty}^\infty dq\langle E+\hbar A_l(E)|q\rangle
{\hbar\over i}{\partial\over\partial q}\langle q| E\rangle
\nonumber\\
&&\approx
{1\over 2\pi i A_l(E)}\oint dq \left[p(E,q)+{\hbar\over 2i}{1\over p(E,q)}
{\partial p(E,q)\over\partial q}-\hbar^2 {\partial S_2(E,q)\over
\partial q}\right]\\
&&\times {\exp\left({i\over\hbar}\left\{S_0(E,q)-S_0(E+\hbar A_l(E),q)-\hbar^2
\left[S_2(E,q)-S_2(E+\hbar A_l(E),q)\right]\right\}\right)
\over\sqrt{C(E)p(E,q)C(E+\hbar A_l(E))p(E+\hbar A_l(E),q)}}\nonumber.
\end{eqnarray}
A systematic expansion in terms of $\hbar$ then leads to
\begin{equation}
Q_l(E)=Q_l^{(0)}(E)+\hbar Q_l^{(1)}(E)+{1\over 2}\hbar^2 Q_l^{(2)}(E)
+{1\over 6}\hbar^3 Q_l^{(3)}(E)+{\cal O}(\hbar^4)
\end{equation}
where the zeroth order term is given by
\begin{equation}
Q_l^{(0)}(E)={1\over 2\pi i l}\oint dq \exp\left[-il\omega(E)t(E,q)
\right].
\end{equation}
Here we have introduced the time
\begin{equation}
t(E,q):={\partial S_0(E,q)\over\partial E}
\end{equation}
spent by a particle of energy $E$ on its
way from the turning point to $q$. Transforming the integration
variable and integrating by parts, it is seen that
\begin{equation}
Q_l^{(0)}(E)={\omega(E)\over 2\pi}\int_0^{2\pi/\omega(E)}dt q(E,t)
e^{-il\omega(E)t}
\end{equation}
is just the Fourier coefficient of the classical trajectory introduced in 
Eq.~(\ref{fourier}). The first order correction is completely
determined by the zeroth order term by means of
\begin{equation}
Q_l^{(1)}(E)={l \omega(E)\over 2}\left[Q_l^{(0)}\right]'(E),
\end{equation}
and the second order term is given by
\begin{eqnarray}
Q_l^{(2)}(E)&=&
{l^2\omega(E)\over 4}\left[\omega Q_l^{(0)}\right]''(E)
-{l^2\over 24}[\omega^2]''(E)Q_l^{(0)}(E)\nonumber\\
&&-2\left[J_2'(E)+C^{(2)}(E)\right]\omega(E)Q_l^{(0)}(E)\label{Q2}\\
&&+{\omega(E)\over 2\pi}\oint dq e^{-il\omega(E)t(E,q)}
\Biggl(2{\partial\over\partial E}S_2(E,q)
-{2i\over l\omega(E)p(E,q)}{\partial\over \partial q}S_2(E,q)\nonumber\\
&&\qquad\qquad-2\omega(E)t(E,q) J_2'(E)
-{7il\omega(E)M^2\over 12 p(E,q)^4}\nonumber\\
&&\qquad\qquad
-{l^2\over 12}\left\{\omega(E)^2{\partial^2\over\partial E^2} t(E,q)
+\left[\omega(E)\omega''(E)-2\omega'(E)^2\right]t(E,q)\right\}
\Biggr)\nonumber.
\end{eqnarray}
In deriving Eq.~(\ref{Q2}) we have made repeated use of partial
integrations.
Note that the third order coefficient $Q_l^{(3)}(E)$, which is also
needed in the sequel, is completely determined by $Q_l^{(0)}(E)$ and
$Q_l^{(2)}(E)$. As a consequence of the symmetry relation
\begin{equation}
\langle n|q|m\rangle=\langle m|q|n\rangle,
\end{equation}
or, equivalently, 
\begin{equation}
Q_l(E)=Q_{-l}(E+\hbar A_l(E)),
\end{equation}
one finds with the help of Eqs.~(\ref{semicl}) and (\ref{wkb})
\begin{equation}
Q_l^{(3)}=\omega\left({3 l\over 2}\left[Q_l^{(2)}\right]'
-3l\omega J_2'\left[Q_l^{(0)}\right]'
+{l^3\over 8}\left\{\left[\omega^2\right]'\left[Q_l^{(0)}\right]'\right\}'
-{l^3\over 4}\left\{\omega^2\left[Q_l^{(0)}\right]'\right\}''\right).
\end{equation}
This way we have expressed the quantum corrections to the dipole matrix
element up to order $\hbar^3$ by purely classical quantities.

For the semiclassical expansion of the master equation we need the
squared matrix element
\begin{equation}
B_l(E):=\langle E+\hbar A_l(E)|q| E\rangle^2\label{squared}.
\end{equation}
From the above results, its semiclassical expansion is given by
\begin{equation}
B_l(E)=B_l^{(0)}(E)+\hbar B_l^{(1)}(E)+{\hbar^2\over 2}B_l^{(2)}(E)
+{\hbar^3\over 6}B_l^{(3)}(E)+{\cal O}(\hbar^4)\label{Bsc}
\end{equation}
where
\begin{eqnarray}
B_l^{(0)}&=&\left[Q_l^{(0)}\right]^2,\nonumber\\
B_l^{(1)}&=&{l\over 2}\omega\left[B_l^{(0)}\right]',
\nonumber\\
B_l^{(2)}&=&{l^2\over 2}\omega^2\left\{\left[Q_l^{(0)}\right]'\right\}^2
+2Q_l^{(0)} Q_l^{(2)},\\
B_l^{(3)}&=&\left\{-{l^3\over 8}\omega \left[\omega^2\right]''
+3 l \omega^2 J_2'\right\} \left[B_l^{(0)}\right]'-{l^3\over 4}
\left\{\omega^3\left[B_l^{(0)}\right]''\right\}'
+{3l\over 2}\omega \left[B_l^{(2)}\right]',\nonumber
\end{eqnarray}
where we have again suppressed the explicit energy dependence.
\subsection{Semiclassical expansion of the probability}
Let us now assume that there is a continuous function $p_{\rm sc}(x)$ 
which interpolates between the discrete values of $p_n$ such that
\begin{equation}
p_{\rm sc}(n)=p_n, \quad\mbox{for}\ n\ \mbox{integer},
\end{equation}
as it is the case for the stationary solution (\ref{stationary}) of
Eq.~(\ref{master}). The probability (\ref{def}) can now be written as
\begin{equation}
P(E)=\sum_{n=0}^\infty\delta(E-E_{\rm sc}(n))p_{\rm sc}(n)=
P_{\rm sc}(E)\sum_{n=0}^\infty\delta(N_{\rm sc}(E)-n)\label{pPsc},
\end{equation}
where we have introduced
\begin{equation}
P_{\rm sc}(E)=\rho_{\rm sc}(E) p_{\rm sc}(N_{\rm sc}(E)),
\end{equation}
and have made use of Eq.~(\ref{wkb}) and the fact that 
\begin{equation}
\delta(E-E_{\rm sc}(n))=\rho_{\rm sc}(E)\delta(N_{\rm sc}(E)-n),
\end{equation}
which is a version of the well-known relation
\begin{equation}
\delta(x)=|f'(x_0)|\delta(f(x))
\end{equation}
holding for functions $f(x)$ with a single zero $x_0$.
Now, the semiclassical expansion of $P(E)$ can be obtained in the
following way. With the help of the identity
\begin{equation}
\sum_{n=0}^\infty \delta(x-n)
=\delta(x)+e^{-{\partial\over\partial x}}\sum_{n=0}^\infty \delta(x-n)
\end{equation}
we can write formally
\begin{equation}
\sum_{n=0}^\infty \delta(x-n)
={{1\over 2}{\partial\over\partial x}\over\sinh({1\over 2}
{\partial\over\partial x})}\Theta(x+{1\over 2}).
\end{equation}
Here the rhs is merely an abbreviation for the series
\begin{equation}
{x\over\sinh(x)}=\sum_{n=0}^\infty {2-2^{2n}\over (2n)!}B_{2n}
x^{2n}=1-{1\over 6}x^2+{7\over 360}x^4+\dots
\end{equation}
where the $B_n$ are Bernoulli numbers.
Now, in view of
\begin{equation}
{\partial\over\partial N_{\rm sc}}={\partial E\over\partial N_{\rm sc}}
{\partial\over\partial E}
\end{equation}
we have
\begin{eqnarray}
{\rho(E)\over\rho_{\rm sc}(E)}=\sum_{n=0}^\infty \delta(N_{\rm sc}(E)-n)
&=&{{1\over 2\rho_{\rm sc}(E)}{\partial\over\partial E} \over\sinh(
{1\over 2\rho_{\rm sc}(E)}{\partial\over\partial E})}
\Theta(N_{\rm sc}(E)+{1\over 2})\nonumber\\
&=&{{\hbar\Omega(E)\over 2}{\partial\over\partial E} \over\sinh(
{\hbar\Omega(E)\over 2}{\partial\over\partial E})}
\Theta(N_{\rm sc}(E)+{1\over 2}).
\label{rhrhsc}
\end{eqnarray}
From Eq.~(\ref{wkb}) we find
\begin{equation}
\Theta(N_{\rm sc}(E)+{1\over 2})=\Theta(E)-\hbar^2\omega(0)J_2(0)
\delta(E)+{\cal O}(\hbar^3),
\end{equation}
where we have used the fact that $J_0(0)=0$. This combines with
Eqs.~(\ref{rhrhsc}) and (\ref{coarse}) to yield the semiclassical
expansion
\begin{equation}
{\rho(E)\over\rho_{\rm sc}(E)}=\Theta(E)-{\hbar^2\over 24}\omega(E)
\omega(0)\delta'(E)-\hbar^2\omega(0)J_2(0)\delta(E)+{\cal O}(\hbar^4).
\end{equation}
Hence, from Eq.~(\ref{pPsc}), the probability reads in the
semiclassical limit
\begin{equation}
P(E)=P_{\rm sc}(E)\left[\Theta(E)-{\hbar^2\over 24}\omega(E)
\omega(0)\delta'(E)-\hbar^2\omega(0)J_2(0)\delta(E)
+{\cal O}(\hbar^4)\right].
\label{PPsc}
\end{equation}
Now, as a consequence of Eq.~(\ref{stationary}), we have
\begin{equation}
P_{\rm sc}^{({\rm eq})}(E)={e^{-\beta E}\over Z}\rho_{\rm sc}(E).
\end{equation}
The equilibrium distribution function then takes the form
\begin{equation}
P_{\rm eq}(E)={e^{-\beta E}\over \hbar Z}\left\{
{\Theta(E)\over\omega(E)}-\hbar^2\left[(J_2\Theta)'(E)
+{\omega(0)\over 24}\delta'(E)\right]+{\cal O}(\hbar^4)\right\}.
\label{equilibrium}
\end{equation}
Thus, to second order in $\hbar$ equilibrium expectation values
(\ref{expectation}) are given by
\begin{eqnarray}
\langle f(E)\rangle_{\rm eq}&=&{1\over\hbar Z}\Biggl\{
\int_0^\infty dE e^{-\beta E}
\left[{1\over\omega(E)}-\hbar^2J_2'(E)\right]f(E)\nonumber\\
&&+{\hbar^2\over 24}
\omega(0)\left[f'(0)-\beta f(0)\right]-\hbar^2J_2(0)f(0)
+{\cal O}(\hbar^4)\Biggr\}.\label{erw}
\end{eqnarray}
Using this for $f(E)\equiv 1$, the partition function $Z$ is found to read
\begin{equation}
\hbar Z =\hbar Z_{\rm cl}-\hbar^2\left[\int_0^\infty dE
J_2'(E)e^{-\beta E}+{1\over 24}\beta\omega(0)+J_2(0)\right]
+{\cal O}(\hbar^4)\label{Zustandss}
\end{equation}
where
\begin{equation}
Z_{\rm cl}={1\over\hbar}\int_0^\infty dE {e^{-\beta E}\over\omega(E)}.
\end{equation}
is the classical partition function. Of course, truncation of the
expansion in terms of $\hbar$ limits the range of temperatures to
$k_B T\gg \hbar\omega_0$ as is readily seen from
Eqs.~(\ref{equilibrium})-(\ref{Zustandss}).
\subsection{Semiclassical expansion of the transition rate}
To expand the transition rate in the semiclassical limit we define
\begin{equation}
W_l(E):=W_{N_{\rm sc}(E)+l,N_{\rm sc}(E)}
\label{defWl}
\end{equation}
which reads in view of Eqs.~(\ref{transi}) and (\ref{squared})
\begin{equation}
W_l(E)={1\over\hbar^2}B_l(E)D(\hbar A_l(E)).
\end{equation}
The semiclassical expansion is given by
\begin{equation}
W_l(E)={1\over\hbar^2}W_l^{(0)}(E)+{1\over\hbar}W_l^{(1)}(E)
+W_l^{(2)}(E)+\hbar W_l^{(3)}(E)+\dots\label{Wlsc}
\end{equation}
Introducing the cosine moment of the damping kernel
\begin{equation}
\gamma_c(z)={\rm Re}\, \hat{\gamma}(iz),
\end{equation}
the coefficients of the expansion (\ref{Wlsc}) take the form
\begin{eqnarray}
W_l^{(0)}&=&{2M\over\beta} B_l^{(0)}\gamma_c,\nonumber\\
W_l^{(1)}&=&{lM\omega\over\beta}\left[\left(-\beta B_l^{(0)}+
\left[B_l^{(0)}\right]'\right)\gamma_c+|l|\omega'B_l^{(0)}
\gamma_c'\right],\nonumber\\
W_l^{(2)}&=&{M\over12\beta} \Biggl[\left(l^2\left\{
- 6\beta\omega\omega'
+ 2\beta^2 \omega^2 \right\}B_l^{(0)}
- 6\beta l^2\omega^2 \left[B_l^{(0)}\right]'
+ 12B_l^{(2)}\right)\gamma_c\nonumber\\&&
+ \left(\left\{|l|^3\left[4\omega^2\omega''
+ 4\omega(\omega')^2
- 6\beta\omega^2\omega'\right]
+ 24|l|\omega^2 J_2'\right\}B_l^{(0)}
+ 6|l|^3\omega^2\omega'\left[B_l^{(0)}\right]'\right)\gamma_c'\\&&
+ 3l^4 \omega^2(\omega')^2 B_l^{(0)}\gamma_c''\Biggr],\nonumber\\
W_l^{(3)}&=&{lM\omega\over 24\beta}\Biggl[\biggl(\left\{l^2\left[
- 4 \beta (\omega')^2 
- 4 \beta \omega \omega'' 
+ 4 \beta^2 \omega \omega'\right]
- 24 \beta \omega J_2'\right\} B_l^{(0)}\nonumber\\&&
+\left\{l^2\left[
- 2 \omega \omega'' 
- 2 (\omega')^2 
- 6 \beta \omega \omega' 
+ 2 \beta^2 \omega^2\right]
+ 24 \omega J_2' \right\}\left[B_l^{(0)}\right]' \nonumber\\&&
- 6 l^2 \omega \omega' \left[B_l^{(0)}\right]'' 
- 2 l^2 \omega^2 \left[B_l^{(0)}\right]''' 
- 12 \beta B_l^{(2)}
+ 12 \left[B_l^{(2)}\right]'\biggr)\gamma_c\nonumber\\&&
+\biggl(\biggl\{|l|^3\left[
  2  \omega^2 \omega''' 
+ 2  (\omega')^3 
+ 8  \omega \omega' \omega''
- 10 \beta \omega (\omega')^2
- 4  \beta \omega^2 \omega'' 
+ 2  \beta^2 \omega^2 \omega' \right]\nonumber\\&& 
+ |l|\left[72 \omega \omega' J_2' 
+ 24  \omega^2 J_2''
- 24  \beta \omega^2 J_2'\right] \biggr\}B_l^{(0)}\nonumber\\&&
+ \left\{|l|^3\left[4 \omega (\omega')^2
+ 4 \omega^2 \omega'' 
- 6 \beta \omega^2 \omega'\right] 
+ 24 |l|\omega^2 J_2'\right\}\left[B_l^{(0)}\right]' 
+ 12 |l| \omega' B_l^{(2)} \biggr)\gamma_c'\nonumber\\&&
+\biggl(\left\{l^4\left[
  4 \omega (\omega')^3 
+ 4 \omega^2 \omega'\omega''
- 3 \beta \omega^2 (\omega')^2\right]
+ 24 l^2 \omega^2 \omega' J_2' \right\}B_l^{(0)}
+ 3 l^4 \omega^2 (\omega')^2 \left[B_l^{(0)}\right]' 
\biggr)\gamma_c''\nonumber\\&&
+ |l|^5 \omega^2 (\omega')^3 B_l^{(0)} \gamma_c'''
\Biggr],\nonumber
\end{eqnarray}
where $\gamma_c^{(i)}(l\omega(E))$ was abbreviated by $\gamma_c^{(i)}$
and the energy dependence was again suppressed.
\subsection{Semiclassical expansion of the master equation}
We are now prepared to address the expansion of the master equation
(\ref{zwischen}), which may be rewritten as
\begin{equation}
\dot{P}(E)=\sum_{n=0}^\infty\sum_{l=-n}^\infty 
[\delta(E-E_n-\hbar A_l(E_n))-\delta(E-E_n)]W_l(E_n)p_n\label{zwischen1}.
\end{equation}
Using the relations
\begin{equation}
\left[\delta(E-E_n-\hbar A_l(E_n))-\delta(E-E_n)\right]f(E_n)
=\sum_{k=1}^\infty \left({\partial\over\partial E}\right)^k{[-\hbar
A_l(E)]^k\over k!}\delta(E-E_n)f(E)
\end{equation}
and
\begin{equation}
\sum_{n=0}^\infty\sum_{l=-n\atop l\neq 0}^\infty f(n,l)=\sum_{l=1}^\infty\left\{
\sum_{n=0}^\infty [f(n,l)+f(n,-l)]- \sum_{n=0}^{l-1}f(n,-l)\right\},
\end{equation}
we find from Eq.~(\ref{zwischen1})
\begin{equation}
\dot{P}(E)=\sum_{k=1}^\infty \left({\partial\over\partial E}\right)^k
{(-\hbar)^k\over k!}
\sum_{l=1}^\infty\left\{[A_l(E)]^kW_l(E)+[A_{-l}(E)]^kW_{-l}(E)\right\}
P(E)+P_B(E)
\label{zwischen2}
\end{equation}
where 
\begin{equation}
P_B(E):=-\sum_{k=1}^\infty \left({\partial\over\partial E}\right)^k
{(-\hbar)^k\over k!}\sum_{l=1}^\infty
[A_{-l}(E)]^kW_{-l}(E)\sum_{n=0}^{l-1}\delta(E-E_n)p_n\label{bound}
\end{equation}
is a boundary term which vanishes since
\begin{equation}
W_{-l}(E_n)\propto B_{-l}(E_n)=\langle n-l|q|n\rangle=0 \quad
\mbox{for}\ n\leq l-1.
\end{equation}

Despite the fact that the expansion (\ref{Wlsc})
of $W_l(E)$ starts with a term of order $\hbar^{-2}$, the series
expansion in $\hbar$ of the rhs of Eq.~(\ref{zwischen2}) does not
contain terms of order $\hbar^{-1}$ since they vanish due to the
symmetry of $\gamma_c(l\omega(E))$ and $Q_l^{(0)}(E)$ in $l$. 
In leading order we then have 
\begin{equation}
\dot{P}(E)={\cal L}_0 P(E)
\end{equation}
where
\begin{equation}
{\cal L}_0={1\over\beta}{\partial\over\partial E}e^{-\beta E}\Lambda^{(0)}(E)
{\partial\over\partial E}e^{\beta E}{\omega(E)\over 2\pi}
\label{L0}
\end{equation}
with $\Lambda^{(0)}(E)$ defined in Eq.~(\ref{Lambda}). Hence, we have
recovered the Zwanzig energy diffusion equation (\ref{zwanzig}).

The next order correction of order $\hbar$ vanishes by
symmetry. Thus, the leading quantum corrections are of order
$\hbar^2$. Inserting the expansions provided in the previous
subsections, one obtains a large number of terms that, after some
manipulations, may be combined to give
\begin{equation}
\dot{P}(E)=({\cal L}_0+\hbar^2 {\cal L}_2) P(E)\label{enderg}
\end{equation}
where
\begin{eqnarray}
{\cal L}_2&=&{1\over\beta}{\partial \over\partial E}e^{-\beta E}
\left[\Lambda^{(2)}(E)+\Phi(E)+{\partial\over\partial E}\chi(E)e^{\beta E}
{\partial\over\partial E}e^{-\beta E}\right]
{\partial\over\partial E}e^{\beta E}{\omega(E)\over 2\pi}\nonumber\\
&&+{\cal L}_0 J_2'(E)\omega(E).
\label{L2}
\end{eqnarray}
We also have introduced the functions
\begin{eqnarray}
\chi&=&{\pi M\omega^3\over 3}
\sum_{l=1}^\infty l^4 B_l^{(0)}\gamma_c, \label{chi}\\
\Lambda^{(2)}&=&2\pi M\omega\sum_{l=1}^\infty\left\{
l^2\left[2 \omega J_2'B_l^{(0)}+B_l^{(2)}\right]\gamma_c+
2l^3 \omega^2 J_2' B_l^{(0)} \gamma_c'\right\},\label{Lambda2}
\end{eqnarray}
and
\begin{equation}
\Phi= {\pi M\omega\over 18}\sum_{l=1}^\infty\Biggl(3 l^4 \biggl\{
(\omega^2)'' B_l^{(0)}-2\omega^2\left[B_l^{(0)}\right]''\biggr\}
\gamma_c
+2 l^5 \omega\left[(\omega^3)'B_l^{(0)}\right]'\gamma_c'
+3 l^6 \omega^2(\omega')^2 B_l^{(0)}\gamma_c''\Biggr)\label{phi}.
\end{equation}
Note that the function (\ref{Lambda2}) combines with Eq.~(\ref{Lambda}) to
the generalized energy relaxation coefficient
\begin{equation}
\Lambda(E)=4\pi M\Omega(E)\sum_{l=1}^\infty l^2 B_l(E)\gamma_c(l\Omega(E))
=\Lambda^{(0)}(E)+\hbar^2\Lambda^{(2)}(E)+{\cal O}(\hbar^3).\label{Lambdagen}
\end{equation}

\section{Generalized energy diffusion equation}
The results of the previous section can be summarized as follows.
The generalized energy diffusion equation containing the leading
quantum corrections can be written in the form
\begin{eqnarray}
\dot{P}(E)&=&{1\over\beta}{\partial\over\partial E}
\Biggl\{e^{-\beta E}\Biggl[\Lambda(E)+\hbar^2\Phi(E)
\nonumber\\&&+\hbar^2{\partial\over\partial E}\chi(E)e^{\beta E}
{\partial\over\partial E}e^{-\beta E}\Biggr]
{\partial\over\partial E}e^{\beta E}{\Omega(E)\over 2\pi}
\Biggr\}P(E)={\cal L}P(E)
\label{zwanzop}
\end{eqnarray}
Here $\Omega(E)$ is a generalized oscillation frequency containing
quantum corrections as defined in Eq.~(\ref{defom}). $\Lambda(E)$ is 
given in Eq.~(\ref{Lambdagen}) and contains the obvious quantum
corrections to the classical energy relaxation coefficient. The
functions $\Phi(E)$ and $\chi(E)$ were introduced in Eqs.~(\ref{phi})
and (\ref{chi}), respectively. Apart from the classical quantities
$\omega(E)$, $Q_l^{(0)}(E)$, and $\hat{\gamma}(z)$ that are
ingredients of the classical Zwanzig equation, one merely needs to
calculate $J_2(E)$ given in Eq.~(\ref{defIk}) and the second order
quantum correction $Q_l^{(2)}(E)$ of the dipole matrix element given
in Eq.~(\ref{Q2}) in order to determine the generalized
Fokker-Planck operator in Eq.~(\ref{zwanzop}) explicitly.
Since one can show that the equilibrium solution (\ref{equilibrium})
is a stationary solution of Eq.~(\ref{zwanzop}), the only non-trivial
quantum correction that survives in equilibrium is seen to be the
quantity $J_2(0)$. 

In the case of an Ohmic bath, i.e. $\gamma_c(\omega)=\gamma$, the
coefficients become particularly simple. Given the functions $\omega$,
$J_2$, and $B_l^{(0/2)}$, all quantities entering Eq.~(\ref{zwanzop})
can be expressed in terms of three sums
\begin{eqnarray}
\lambda^{(i)}&=&4\pi M \gamma
\sum_{l=1}^\infty l^2 B_l^{(i)},\ \ (i=0,2)\ ,\nonumber\\
\phi&=&4\pi M \gamma\sum_{l=1}^\infty l^4 B_l^{(0)}. \label{hilfs}
\end{eqnarray}
Then
\begin{eqnarray}
\Omega&=&\omega+\hbar^2\omega^2 J_2',\nonumber\\
\chi&=&{1\over 12}\omega^3\phi,\nonumber\\
\Phi&=&{\omega\over 24}\left[(\omega^2)''\phi-2\omega^2\phi''\right]
,\label{ohmic}\\
\Lambda&=&\omega\lambda^{(0)}+\hbar^2\left[\omega^2 J_2' \lambda^{(0)}
+{1\over 2}\omega\lambda^{(2)}\right]\nonumber.
\end{eqnarray}
\subsection{Harmonic oscillator}
The simplest model for a molecular system, which already exhibits many
of the characteristic features of VER, is the harmonic oscillator
potential: 
\begin{equation}
V(q)={1\over 2}M\omega_0^2 q^2.\label{Vharm}
\end{equation}
For the potential (\ref{Vharm}) one obtains
\begin{eqnarray}
\omega(E)&=&\omega_0,\nonumber\\
J_2(E)&=&0,\\
Q_{\pm1}^{(0)}(E)&=&\sqrt{E\over 2M\omega_0^2},\nonumber
\end{eqnarray}
and the squared matrix elements read
\begin{equation}
B_{\pm 1}(E)={1\over 2M\omega_0^2}\left(E\pm \hbar{\omega_0\over 2}\right).
\end{equation}
Special to the case of the harmonic oscillator is the fact that
$B_l(E)$  vanishes for $l\neq\pm 1$.
The transition rates are
\begin{equation}
W_{\pm 1}(E)=\pm{\gamma_c\over\hbar\omega_0}\bar{n}(\pm\hbar\omega_0)
\left(E\pm{\hbar\omega_0\over 2}\right)
\end{equation}
where $\bar{n}(E)$ was introduced in Eq.~(\ref{barn}).
The Pauli master equation reads in terms of the probability $P(E)$ 
\begin{eqnarray}
\dot{P}(E)&=&{\gamma_c\over\hbar\omega_0}\Biggl[\bar{n}(\hbar\omega_0)\left(
e^{-\hbar\omega_0{\partial\over\partial E}}-1\right)\left(E+{\hbar\omega_0\over
2}\right)\nonumber\\
&&-\bar{n}(-\hbar\omega_0)\left(
e^{\hbar\omega_0{\partial\over\partial E}}-1\right)\left(E-{\hbar\omega_0\over
2}\right)\Biggr]P(E).
\end{eqnarray}
For the energy relaxation coefficient (\ref{Lambda}) and the coefficients
(\ref{chi})-(\ref{phi}) one finds
\begin{eqnarray}
\Lambda^{(0)}(E)&=&{\gamma_c E\over\omega_0},\nonumber\\
\chi(E)&=&{\gamma_c\omega_0 E\over 12},\nonumber\\
\Lambda^{(2)}(E)&=&0,\\
\Phi(E)&=&0,\nonumber
\end{eqnarray}
so that the generalized energy diffusion equation (\ref{zwanzop})
takes the form
\begin{eqnarray}
\dot{P}(E)&=&{\gamma_c\over\beta}{\partial\over\partial E}
\Biggl(E[\beta P(E)+P'(E)]\nonumber\\
&&+{\hbar^2\omega_0^2\over 12}\biggl\{[\beta^2 P'(E)+2\beta P''(E)
+P^{(3)}(E)]E+\beta P'(E)+P''(E)\biggr\}\Biggr).\label{harmP}
\end{eqnarray}
This equation can be solved in terms of the Fourier transform
\begin{equation}
\tilde{P}(u)=\int_{-\infty}^\infty dE e^{iuE}P(E).
\end{equation}
Despite the fact that $P(E)=0$ for $E<0$, the integral runs over all
energies to collect possible contributions of delta functions at
$E=0$. The Fourier transform of Eq.~(\ref{harmP}) takes the form
\begin{eqnarray}
\tilde{P}(u)&=&{i\gamma_c\over\beta}u
\Biggl\{\tilde{P}(u)+i(\beta-iu)\tilde{P}'(u)
\nonumber\\
&&+{\hbar^2\omega_0^2\over 12}(\beta-iu)
\biggl[(\beta-2iu)\tilde{P}(u)+u
(\beta-iu)\tilde{P}'(u)\biggr]\Biggr\}.\label{eqtrans}
\end{eqnarray}
This first order partial differential equation can be solved by the
method of characteristics. The general solution is given by
\begin{eqnarray}
\tilde{P}(u)&=&{\beta \over \beta-i(1-e^{-\gamma_c t})u}
\tilde{P}_0\left({-i\beta e^{-\gamma_c t}u
\over \beta-i(1-e^{-\gamma_c t})u}\right)\nonumber\\
&&+i\hbar^2 {\beta \omega_0^2 (1-e^{-\gamma_c t})u
(\beta-iu)\over 12[\beta-i(1-e^{-\gamma_c t})u]^3}
\Biggl\{{-i\beta^2(\beta-iu)e^{-\gamma_c t}u
\over \beta-i(1-e^{-\gamma_c t})u}
\tilde{P}_0'\left({-i\beta e^{-\gamma_c t}u
\over \beta-i(1-e^{-\gamma_c t})u}\right)
\label{loesharm}\\
&&+{1\over 2} \left[-(1-e^{-\gamma_c t}) u^2-i\beta(3+e^{-\gamma_c t})u
+2\beta^2\right] 
\tilde{P}_0\left({-i\beta e^{-\gamma_c t}u\over 
\beta-i(1-e^{-\gamma_c t})}\right)\Biggr\}\nonumber
\end{eqnarray}
where $\tilde{P}_0(u)$ is the initial distribution.
For $t\to\infty$ this solution tends to
\begin{equation}
\tilde{P}(u)={\beta\over \beta-iu}-{\hbar^2\omega_0^2\beta\over 24}
\left(\beta-iu-{\beta^2\over \beta-iu}\right)
\label{harmeq}
\end{equation}
where we have used $\tilde{P}_0(0)=1$ which is just the
normalization condition. The inverse Fourier transform of this
stationary probability gives 
\begin{equation}
P(E)=\Theta(E)\beta\left(1+{\hbar^2\omega_0^2\beta^2\over
24}\right)e^{-\beta E}-{\hbar^2\omega_0^2\beta\over
24}\left(\beta\delta(E)+\delta'(E)\right).
\end{equation}
This result coincides with Eq.~(\ref{equilibrium}) if one inserts the
quantum mechanical partition function
$Z=1/[2\sinh(\hbar\omega_0\beta/2)]$ and expands up to second order in
$\hbar$.

As a specific example consider a system with given initial energy
$E_0$, i.e.
\begin{equation}
P_0(E)=\delta(E-E_0).
\end{equation}
Then moments are readily calculated from Eq.~(\ref{loesharm}) using
\begin{equation}
\tilde{P}(u)=1+iu\langle E(t)\rangle-{u^2\over 2}\langle E^2(t)\rangle+\dots.
\end{equation}
This way the expectation value is found to read
\begin{equation}
\langle E(t)\rangle=
E_0 e^{-\gamma_c t}+\left({1\over\beta}
+{\hbar^2\omega_0^2\beta\over 12}\right) \left(1-e^{-\gamma_c t}\right)
\label{Eerwharm}
\end{equation}
and the second moment obeys
\begin{eqnarray}
\langle E^2(t)\rangle&=&{1\over\beta^2}\left[
(\beta^2E_0^2-4 \beta E_0+2) e^{-2\gamma_c t}
+4(\beta E_0-1) e^{-\gamma_c t}+2\right]\nonumber\\
&&+{\hbar^2\omega_0^2\over 12}
\left[(-4\beta E_0+7) e^{-2 \gamma_c t}
+4(\beta E_0-2) e^{-\gamma_c t}+1\right].
\end{eqnarray}
From Eq.~(\ref{Eerwharm}) one finds that for the case of linear
dissipation considered here the energy relaxation time
\begin{equation}
T_1={1\over \gamma_c}\label{T1harm}
\end{equation}
is not modified by quantum effects in accordance with the findings by
Bader and Berne \cite{bader}.
\subsection{Morse oscillator}
The fact that the relaxation time is unaffected by quantum effects is
of course a special feature of the simple harmonic potential
(\ref{Vharm}). As an example of a more realistic model for VER we 
consider a damped Morse oscillator. This system plays an important
role in the modeling of, e.g., diatomic molecular systems.
The Hamiltonian of the Morse oscillator reads 
\begin{equation}
H={P^2\over 2 M}+ D_0\left[1-e^{-\kappa(r-r_0)}\right]^2
\end{equation}
where $D_0$ is the dissociation energy, $r_0$ the nuclear distance at
equilibrium, and $\kappa^{-1}$ the range of the molecular potential.
Here, the $P=M dr/d\tau$ is the classical momentum.
In terms of the scaled variables $q=\kappa(r-r_0)$,
$p=P/\sqrt{2D_0M}$, $t=\tau\kappa\sqrt{2D_0/M}$ the Hamiltonian takes
the form
\begin{equation}
H(p,q)={p^2\over 2}+V(q)
\end{equation}
where
\begin{equation}
V(q)={1\over 2}(1-e^{-q})^2.
\end{equation}
Note that now $\hbar$ is dimensionless and stands for
$\hbar\kappa/\sqrt{2D_0M}$. The classical trajectory for a bound
particle with energy $E<1/2$ is found by solving
\begin{equation}
2E-1=\dot{q}^2+e^{-2 q}-2e^{-q}
\end{equation}
with the result
\begin{equation}
q(E,t)=\ln\left[{1+\sqrt{2E}\cos(t\sqrt{1-2E})\over 1-2E}\right].
\end{equation}
Here we have assumed that the particle is at a turning point at $t=0$.
From this result it is seen that the energy dependent frequency is
given by
\begin{equation}
\omega(E)=\sqrt{1-2E}.
\end{equation}
To find the Fourier coefficients we write the trajectory in the form
\begin{equation}
q(E,t)=\ln\left[{1+\sqrt{1-2E}\over 2(1-2E)}\right]+
\ln[1-z_2(E)e^{i\omega(E)t}]+\ln[1-z_2(E)e^{-i\omega(E)t}]\label{traject}
\end{equation}
where 
\begin{equation}
z_{1/2}(E)=-{1\pm\sqrt{1-2E}\over\sqrt{2E}}.
\end{equation}
An expansion of the logarithms gives for $l\neq 0$ the Fourier
coefficients
\begin{equation}
Q_l(E)=-{[z_2(E)]^{|l|}\over |l|}=-{(-1)^{|l|}\over |l|}
\left[{1-\omega(E)\over 1+\omega(E)}\right]^{|l|/2}\label{four}.
\end{equation}
Further, the action variable $J_0(E)$ is obtained by integration of
Eq.~(\ref{J0strich}) as
\begin{equation}
J_0(E)=\int_0^E{dE'\over\omega(E')}=1-\sqrt{1-2E},
\end{equation}
and the second eikonal $S_2(E,q)$ can also be integrated exactly yielding
\begin{equation}
S_2(E,q)={-6E+8E^2+6e^{-2q}E+1-3e^{-q}+3e^{-2q}-e^{-3q}\over
48 E (2E-1+2e^{-q}-e^{-2q})^{3/2}}.
\end{equation}
Therefore, according to Eq.~(\ref{defIk}) we have $J_2(E)=0$.
Similarly, one finds
\begin{equation}
C^{(2)}(E)=0.
\end{equation}
The contour integral in Eq.~(\ref{Q2}) may be
evaluated conveniently by means of the substitutions
\begin{eqnarray}
e^{i\omega(E)t}&\to&z,\nonumber\\
dt&\to&{dz\over iz\omega(E)},\nonumber\\
p(E,t)&\to&{i(z^2-1)(z_2^2-1)z_2 \over (z z_2-1)(z_2-z)(z_2^2+1)},\\
\exp[-q(E,t)]&\to&{(z_2^2-1)^2 z\over(zz_2-1)(z_2-z)(z_2^2+1)}.\nonumber
\end{eqnarray}
One then finds
\begin{equation}
Q_l^{(2)}(E)=Q_l^{(0)}(E) |l| {3\omega^2(E) |l|^3-8\omega^3(E)
l^2-12 |l| E^2+2\omega^3(E)\over 96 E^2\omega^2(E)},
\end{equation}
which gives for the squared matrix element
\begin{equation}
B_l^{(2)}(E)=[Q_l^{(0)}]^2(E){\omega^3(E)(1-4l^2)+3|l|(l^2\omega^2(E)-2E^2)
   \over 12 |l|\omega^2(E) E^2}\label{Bmors}.
\end{equation}
Restricting ourselves to the Ohmic case $\gamma_c(\omega)=\gamma$, one
only needs to calculate the functions (\ref{hilfs}) which read
\begin{eqnarray}
\lambda^{(0)}(E)&=&2\pi\gamma\left[(1-2E)^{-1/2}-1\right],\nonumber\\
\lambda^{(2)}(E)&=&2\pi\gamma (E+2) (1-2E)^{-3/2},\\
\phi(E)&=&2\pi\gamma E(1-2E)^{-3/2}.\nonumber\\
\end{eqnarray}
Using Eq.~(\ref{ohmic}) we then find
\begin{eqnarray}
\chi(E)&=&2\pi\gamma{E\over 12},\nonumber\\
\Phi(E)&=&-2\pi\gamma{E+2\over 4(1-2E)^2},\\
\Lambda^{(2)}(E)&=&2\pi\gamma{E+2\over 4(1-2E)^2}.\nonumber
\end{eqnarray}
Thus, the generalized energy diffusion equation takes the form
\begin{equation}
\dot{P}(E)={\gamma\over\beta}{\partial\over\partial E}e^{-\beta E}
\left[J_0(E)+{\hbar^2\over 12}{\partial\over\partial E}E
e^{\beta E}{\partial\over\partial E}e^{-\beta E}\right]
{\partial\over\partial E}e^{\beta E}
\omega(E)P(E).
\end{equation}
We recall that all quantities are dimensionless, in particular 
the dimensionless energy $E$ is measured in units of $2D_0$ and $\beta=2
D_0/k_B T$.

Let us now investigate how an initial probability
\begin{equation}
P_0(E)=\delta(E-E_0)
\end{equation}
evolves for short times. Using
\begin{equation}
\left.{\partial\over\partial t}\langle f(E(t))\rangle\right|_{t=0}
=\int dE {\cal L} P_0(E)f(E)
\end{equation}
one finds for the initial change of the energy expectation value and
the second moment
\begin{eqnarray}
\left.{\partial\over\partial t}\langle E(t)\rangle\right|_{t=0}
&=&-\gamma\left[\omega J_0 -\left(
{1\over\beta}+{1\over 12}\hbar^2\beta\omega \right)\right]
,\nonumber\\
\left.{\partial\over\partial t}\langle E^2(t)\rangle\right|_{t=0}
&=&-2\gamma\left[\omega J_0 -
\left({\omega J_0 +E_0\over\beta}
+{1\over 12}\hbar^2\omega\gamma(2\beta E_0-3)\right)\right].\label{morsmom}
\end{eqnarray}
It is useful to discuss this result in terms of the relaxation rate
$R(E_0)$ defined by
\begin{equation}
\left.{\partial\over\partial t}\langle
E(t)\rangle\right|_{t=0}=-R(E_0) E_0.\label{defR}
\end{equation}
From Eq.~(\ref{morsmom}) one sees that quantum corrections lower the
relaxation rate since the mean thermal energy is raised by zero point
motion. Usually this effect is accounted for by the definition of the
relaxation time $T_1$ according to
\begin{equation}
{\partial\over\partial t}\langle E(t)\rangle=-{1\over T_1}
\left[\langle E(t)\rangle - \langle E\rangle_{\rm eq}\right].
\label{T1}
\end{equation}
Contrary to the harmonic case $\omega J_0\neq E$, and the
energy relaxation is non-exponential for high energies where
anharmonicities are important. The usual concept of a relaxation time
is thus no longer strictly valid. Eq.~(\ref{T1}) may, however, 
serve in connection with Eq.~(\ref{morsmom}) to define an initial
relaxation rate $1/T_1$. For the classical relaxation time one then
finds for $\beta\gg 1$ (that is the potential is almost harmonic for
thermal energies)
\begin{equation}
T_1^{\rm cl}= {E_0-1/\beta\over\gamma(\omega J_0-1/\beta)},\label{T1cl}
\end{equation}
from where it is readily seen that the classical relaxation time
increases as the initial energy increases, consistent with the fact
that the period increases with energy.
The semiclassical corrections to the relaxation time are then found
by using Eqs.~(\ref{erw}) and (\ref{Zustandss}). The leading order
correction to Eq.~(\ref{T1cl}) is found to read
\begin{equation}
{T_1^{\rm sc}\over T_1^{\rm cl}}=1-{1\over 12}\hbar^2 \beta 
\left({1\over E_0}-{1\over J_0}\right)
\end{equation}
from where the classical and harmonic limits are readily recovered by
setting $\hbar=0$ and $J_0(E_0)=E_0$, respectively. With the rates
defined in this way, we now find an acceleration of the relaxation 
process by quantum fluctuations.

\begin{figure}[hbtp]
\psfig{figure=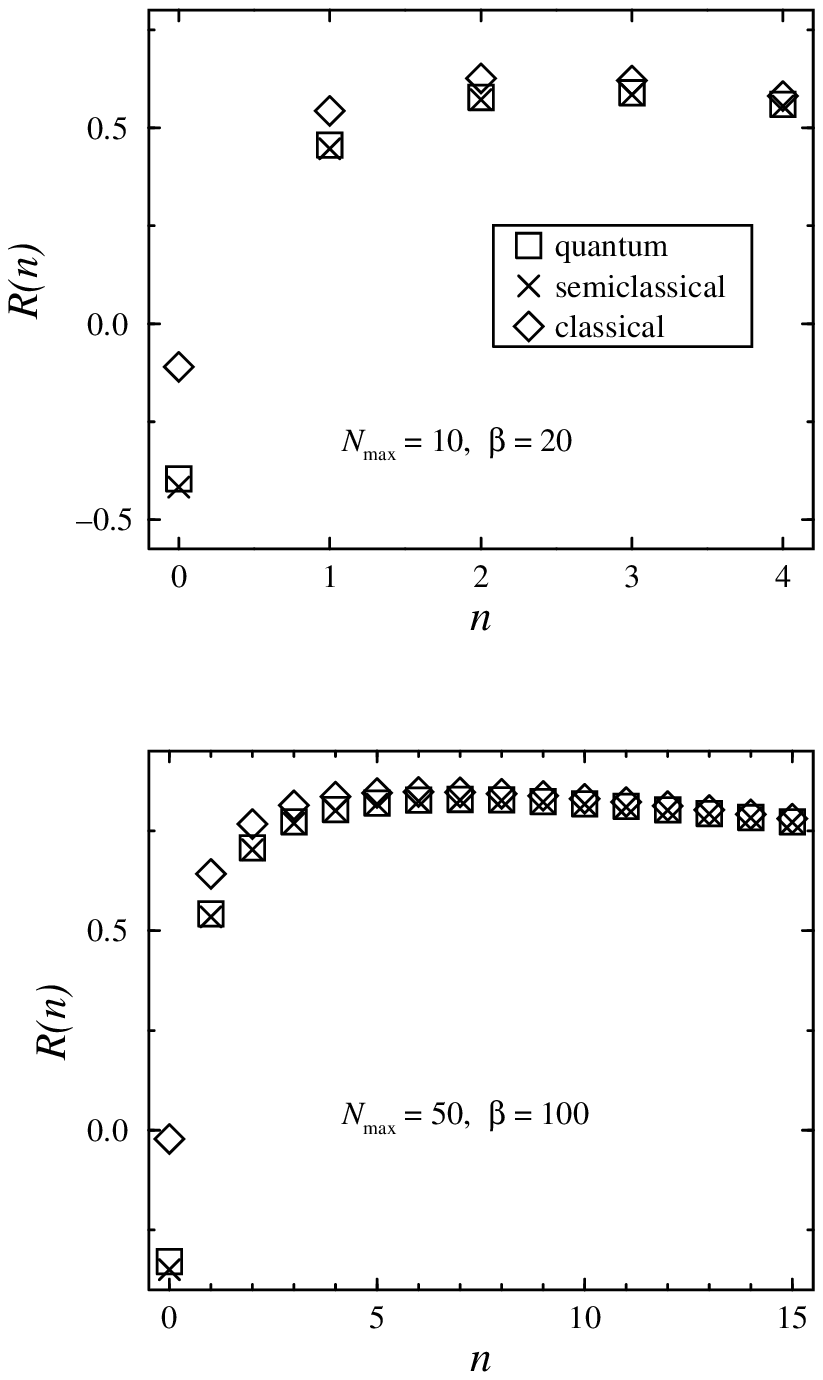}
\caption{Comparison of energy relaxation rates $R(n)$ for various
initial states $n$ based on classical, semiclassical and quantum
theories for the Morse potential and two sets of parameters. $R(n)$ is
given in units of $\gamma_c$. Ground state rates are negative since in
equilibrium also higher vibrational levels are populated.}
\label{zw1}
\end{figure}

To investigate the quality of the semiclassical energy diffusion
equation in some detail, we compare with results of the full master
equation. Since  for the Morse potential exact dipole
matrix elements are available \cite{dipole}, the initial energy
relaxation rate may be calculated using
\begin{equation}
{\partial \over \partial t}
\langle E(t)\rangle =\sum_{n,m=0}^\infty p_m(t) W_{mn}(E_n-E_m)
\label{enerex}
\end{equation}
which is a consequence of the master equation (\ref{master}).
To compare this with the approximation (\ref{morsmom}), one
has to consider an initial distribution
\begin{equation}
p_n(0)=\delta_{n,n_0}.
\end{equation}
For Ohmic damping we then have 
\begin{equation}
\left.{\partial\over\partial t}\langle E(t)\rangle\right|_{t=0}
=-R(n_0)E_0=2\gamma_c \sum_{l=-n_0}^{N_{\rm max}-n_0}
{B_l(E_0)[A_l(E_0)]^2\over e^{\hbar\beta A_l(E_0)}-1}
\end{equation}
where $E_0=E_{n_0}$, 
\begin{equation}
A_l(E)=l\omega(E)-{l^2\over 2}\hbar
\end{equation}
and $B_l(E)$ is given by \cite{dipole}
\begin{equation}
B_l(E)=[Q_l(E)]^2={\left({\omega(E)\over\hbar}-l\right)
\Gamma\left({1\over\hbar}+{1\over 2}+{\omega(E)\over\hbar}\right)
\Gamma\left({1\over\hbar}+{1\over 2}-{\omega(E)\over\hbar}\right)
\over\left({\omega(E)\over\hbar}-{l\over 2}\right)
\Gamma\left({1\over\hbar}+{1\over 2}+{\omega(E)\over\hbar}-l\right)
\Gamma\left({1\over\hbar}+{1\over 2}-{\omega(E)\over\hbar}+l\right)}.
\end{equation}
Figure \ref{zw1} compares the relaxation rates $R(n_0)$ for the
parameters $\hbar=0.0943$, $\beta=20$ (ten bound states) and
$\hbar=0.0198$, $\beta=100$ (50 bound states). Since $N_{\rm
sc}(\langle E\rangle_{\rm eq})=0.27$ and $0.17$, respectively, in
equilibrium only the ground state is populated substantially. Hence,
$R(n)$ is negative for $n=0$ only. The figure shows that for higher
states the classical rates agree well with the quantum results. For
low-lying states, however, while the classical approximation fails
the semiclassical rates are still in good agreement with the master
equation prediction. 
\section{Conclusions}
We have given a systematic semiclassical theory of vibrational energy
relaxation based on the quantum master equation. Specifically, we
have studied a model leading to a generalized Langevin equation in the
classical limit, which is the basis of much of the earlier work. An
expansion in powers of $\hbar$ (more specifically, in powers of
$\hbar\omega_0/k_B T$, where $\hbar\omega_0/2$ is the vibrational zero
point energy) was shown to yield the Zwanzig energy diffusion equation
and quantum corrections thereof. For a harmonic potential, the energy
relaxation time is not affected by quantum fluctuations while it can be
reduced for nonlinear potentials. For the Morse potential the improved
energy diffusion equation was shown to incorporate the main quantum
effects.
\section*{acknowledgments}
The authors would like to thank D.~Richards for helpful correspondence
and acknowledge useful discussions with F.~J.~Weiper and C.~A.~Stafford.
Financial support was provided by the Sonderforschungsbereich 276 of
the Deutsche Forschungsgemeinschaft (DFG, Bonn).

\end{document}